\documentclass[superscriptaddress,amsmath,amssymb, aps, prl,longbibliography,twocolumn, footinbib]{revtex4-2}

\usepackage{mathtools}  

\usepackage{upgreek}
\usepackage{color}
\usepackage{graphicx}
\usepackage{nccmath}
\usepackage{bm}
\usepackage{hyperref}
\hypersetup{colorlinks,breaklinks,
            urlcolor=[rgb]{0,0,0.64},
            linkcolor=[rgb]{0,0,0.64},
            citecolor=[rgb]{0,0,0.64},
            filecolor=[rgb]{0,0,0.64}}
\usepackage{dcolumn}

\newcommand{\beginsupplement} {
    \setcounter{table}{0}
    \renewcommand{\thetable}{S\arabic{table}}
    \setcounter{figure}{0}
    \renewcommand{\thefigure}{S\arabic{figure}}
    \setcounter{equation}{0}
    \renewcommand{\theequation}{S\arabic{equation}}
}

\newcommand{\Harvard}{Department of Physics, Harvard University, Cambridge, Massachusetts 02138, USA.}

\newcommand{\HarvardAppl}{John A. Paulson School of Engineering and Applied Sciences, Harvard University, Cambridge, MA 02138.}

\newcommand{\MaxP}{Max Planck Institute for the Structure and Dynamics of Matter, Luruper Chaussee 149, 22761 Hamburg, Germany.}

\newcommand{\Oxford}{Clarendon Laboratory, University of Oxford, Parks Road, Oxford OX1 3PU, UK.}

\newcommand{\ETH}{Institute for Theoretical Physics, ETH Zurich, 8093 Zurich, Switzerland.}

\begin{document}

\title{ Theory for Anomalous Terahertz Emission in Striped Cuprate Superconductors}

\author{Pavel~E.~Dolgirev}
\thanks{P.E.D. and M.H.M. contributed equally to this work.}
\affiliation{\Harvard}
\email{p\_dolgirev@g.harvard.edu}

\author{Marios~H.~Michael}
\thanks{P.E.D. and M.H.M. contributed equally to this work.}
\affiliation{\Harvard}
\email{marios\_michael@g.harvard.edu}

\author{Jonathan~B.~Curtis}
\affiliation{\Harvard}
\affiliation{\HarvardAppl}

\author{Daniele~Nicoletti}
\affiliation{\MaxP}
\author{Michele~Buzzi}
\affiliation{\MaxP}
\author{Michael~Fechner}
\affiliation{\MaxP}

\author{Andrea Cavalleri}
\affiliation{\MaxP}
\affiliation{\Oxford}
\author{Eugene~Demler}
\affiliation{\ETH}

\date{\today}

\begin{abstract}
Recent experiments in the doped cuprates La$_{2-x}$Ba$_x$CuO$_4$ have revealed the emission of anomalous terahertz radiation after impulsive optical excitation. Here, we theoretically investigate the nonlinear electrodynamics of such striped superconductors and explore the origin of the observed radiation. We argue that photoexcitation is converted into a 
photocurrent
by a second-order optical nonlinearity, which is activated by the breaking of inversion symmetry in certain stripe configurations.  We point out the importance of including Umklapp photocurrents 
modulated at the stripe periodicity itself, which impulsively drive surface Josephson plasmons and lead to a resonant structure of outgoing radiation, consistent with the experiments. We speculate on the utility of the proposed mechanism in the context of generating tunable terahertz radiation.
\end{abstract}

\maketitle

Electronic phases of quantum matter are typically distinguished by their signature electromagnetic responses~\cite{ScalapinoWhiteZhang,Kohn.1964,Yang.1962,TKNN,Qi.2008,Kapitulnik.2009,Imada.1998,Orenstein.2021}. 
For instance, in strongly anisotropic materials, the onset of superconductivity manifests as the appearance of Josephson plasmon (JP) edges in the terahertz reflectivity~\cite{Tamasaku.1992,Uchida.1996,Motohashi.2000,Dienst.2013}. Likewise, states with an incommensurate charge density wave (CDW) exhibit a characteristic quasiparticle gap in the low-frequency conductivity, accompanied by resonance at the pinning frequency~\cite{Lee.1973,Lee.1979,Lee.1993}.
Recent experiment~\cite{stripes1} in the doped single-layer cuprates La$_{2-x}$Ba$_x$CuO$_4$ demonstrated a unique nonlinear optical signature of the ``superstripe'' phase, i.e., when both superconducting and stripe orders are present and intertwined~\cite{Berg_2009,Berg.2009.b,Berg.2009.c,Cremin.2021,Lee.2021,Comin.2016}. The discovered phenomenon can be summarized as follows: upon a strong optical pump pulse, outgoing radiation is observed at terahertz frequencies, with a spectrum peaked at the Josephson plasmon resonance (JPR). These findings are surprising because these materials are nominally centrosymmetric, and such radiation is not expected in the absence of a current or magnetic field bias. In this paper, we provide a theoretical interpretation of these experimental observations. Crucially, this nonlinear effect cannot be understood from the perspective of either the superconducting or stripe orders individually and requires combining both types of symmetry breaking simultaneously.  

The phenomenon of ``radiating stripes'' in LBCO provides several important insights into the nature of this striped superconductor. The presence of a subharmonic optical response constrains the symmetries of the stripe order~\cite{Comin.2015} and, in particular, implies inversion symmetry is broken~\cite{dolgirev2021Umklapp,Kaneko.2021,deLaTorre.2021,Zhao.2017,pettine2023ultrafast}. Consequently, the impulsive optical excitation is now coupled to both bulk and surface Josephson plasmons, which are otherwise symmetry-odd infrared-active modes~\cite{Note2}. We argue that the outgoing terahertz radiation originates from the surface Josephson plasmons and is sensitive to the stripe order, which outcouples these otherwise silent modes. From the observation that the terahertz radiation is sharply peaked at the Josephson plasmon resonance, we can infer that a photocurrent is being generated both at zero momentum and at momenta corresponding to the reciprocal vectors of the CDW lattice, yielding insight into the microscopic physics of the striped state.

\begin{figure}[t!]
\centering
\includegraphics[width=1\linewidth]{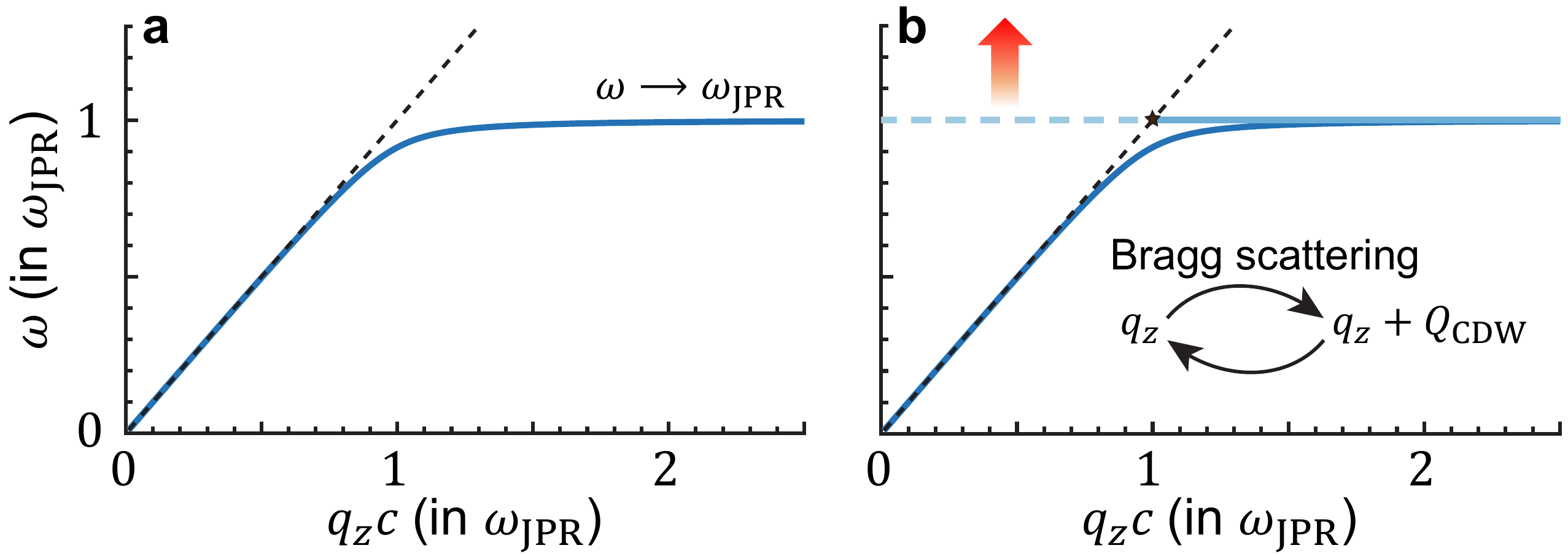} 
\caption{(a) Dispersion $\omega_s(q_z)$ of the surface Josephson plasmons. In strongly anisotropic superconductors, $\omega_s$ quickly saturates at $\omega_{\rm JPR}$. Dashed line corresponds to the light cone $\omega = c q_z$. (b) Schematic of the surface plasmon spectrum in the presence of a weak stripes order, which couples momenta $q_z$ to $q_z + Q_{\rm CDW}$. As such, the plasmon dispersion exhibits back folding to the reduced Brillouin zone, defined by the CDW wave vector $Q_{\rm CDW}$. Notably, the surface plasmons that back fold inside the light cone can now radiate out.}
\label{fig::Plasmon_disp}
\end{figure}

\begin{figure}[t!]
\centering
\includegraphics[width=1\linewidth]{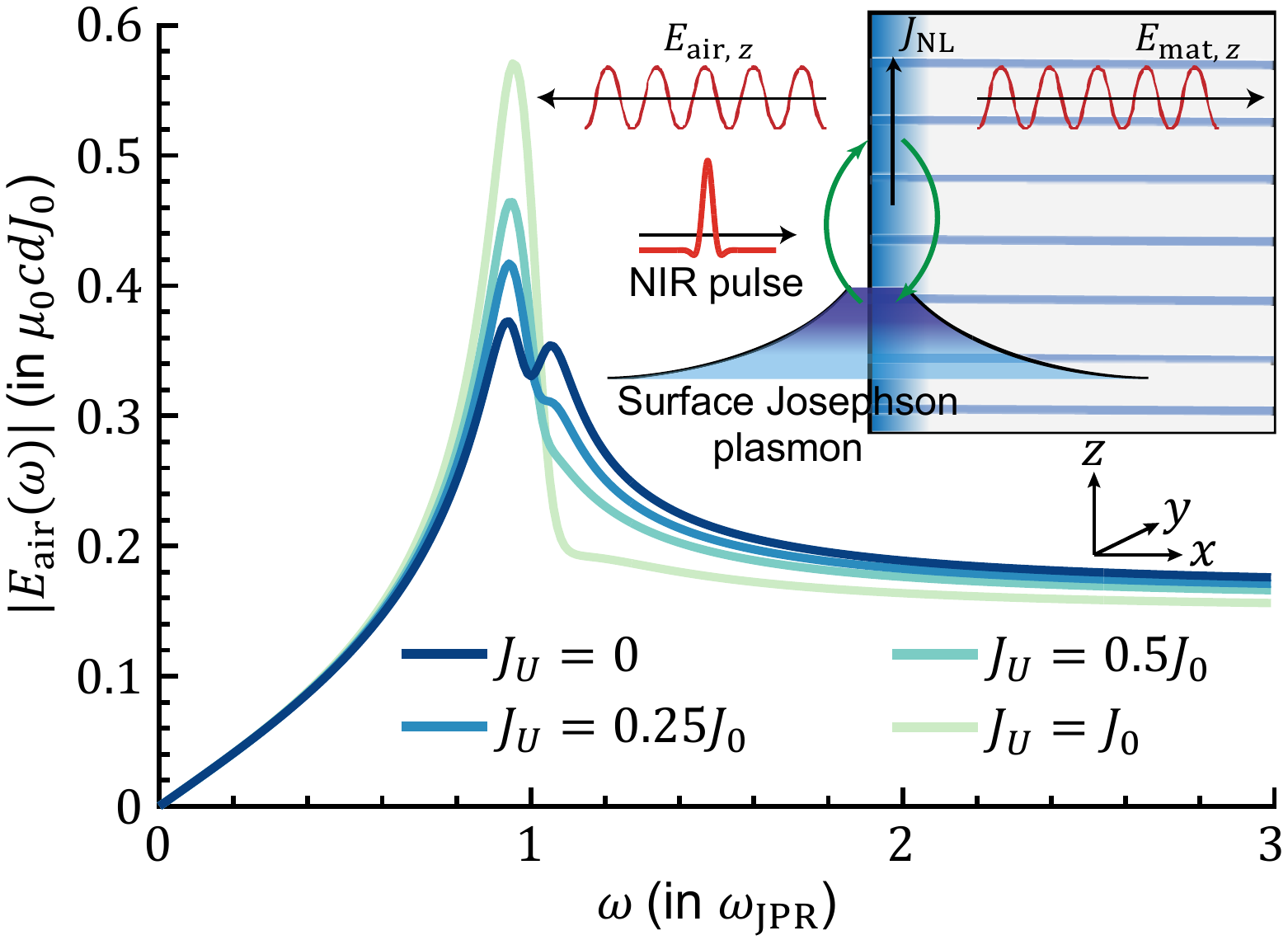} 
\caption{ Emission spectrum.
We find that the outgoing radiation can be sharply peaked around $\omega_{\rm JPR}$ only in the presence of a nonzero Umklapp photocurrent $J_U\neq 0$, Eq.~\eqref{eqn:J_shift}. This Umklapp component impulsively drives high momenta surface Josephson plasmons that can emit light, Fig.~\ref{fig::Plasmon_disp}(b).  
Here we fixed $A = 0.1\varepsilon_{\infty,c}$. 
} 
\label{fig::Filtering_stripes}
\end{figure}

The key finding of this paper can be summarized as follows. The role of the CDW order is that it i) makes the surface modes optically active [Fig.~\ref{fig::Plasmon_disp}] and ii) gives rise to a nonzero Umklapp photocurrent~\cite{dolgirev2021Umklapp}.
We show that Umklapp photocurrents resonantly drive the surface plasmons, which exhibit a large density of states near the Josephson plasma resonance $\omega_{\rm JPR} \simeq 0.5\,$THz, thereby resulting in sharp in frequency radiation [Fig.~\ref{fig::Filtering_stripes}] consistent with the experiment of Ref.~\cite{stripes1}. We point out that the proposed mechanism is generic and potentially useful for designing platforms for the generation of tunable terahertz radiation~\cite{smith1953visible,mikhailov1998plasma,kachorovskii2012current,petrov2016plasma,petrov2017amplified}. We remark that Umklapp currents carry a large momentum of the CDW reciprocal lattice so that nominally, the resulting emitted light is expected to be far away from the light cone and, thus,  decay on a length scale of the order of a few wavelengths at most. Therefore, the proposed mechanism of resonant coupling to  the surface collective modes provides a pathway towards detecting Umklapp currents with far-field optics.

{\it Photocurrent generation}\textemdash A crucial aspect of the experimental findings reported in Ref.~\cite{stripes1} is that the frequency of the pump pulse $\Omega_{\rm pm} \simeq 375\,$THz is at least two orders of magnitude larger than $\omega_{\rm JPR}$. This separation of energy scales suggests that the pump pulse drives the mobile electronic degrees of freedom that downconvert the large incoming frequency into a low-frequency photocurrent~\cite{Aversa1995,Sipe2000,cook2017design,Parker2019Diagrammatic,Xu.2019a,Patankar.2018,Orenstein.2021}. This is further bolstered by the expectation that inversion symmetry is broken for a realistic pattern of charge order~\cite{dolgirev2021Umklapp}. The photocurrent
is characterized by a rank-three conductivity tensor:
\begin{align}
        J_{\rm NL}^a & (\omega)  = \int \frac{d \omega^\prime}{2\pi} \sigma^{(2)}_{abb}(-\omega;\omega^\prime,\omega - \omega^\prime) E^b_{\omega^\prime} E^b_{\omega - \omega^\prime} \label{eqn:shift_current_gen},
\end{align}
where indices $a,b$ represent the Cartesian directions. Following Ref.~\cite{stripes1}, we assume that the
pump-pulse width $\Delta \Omega_{\rm pm}\simeq 5\,$THz is also much larger than $\omega_{\rm JPR}$, 
in turn implying that the photocurrent
appears featureless for $\omega \lesssim \Delta \Omega_{\rm pm}, \Omega_{\rm pm}$:
\begin{align*}
    J_{\rm NL}^a (\omega) \xrightarrow{\omega \lesssim \Delta \Omega_{\rm pm},\Omega_{\rm pm}}  \int \frac{d \omega^\prime}{2\pi} \sigma^{(2)}_{abb}(0;\omega^\prime,- \omega^\prime) E^b_{\omega^\prime} E^b_{- \omega^\prime}.
\end{align*}
This should be contrasted to the experiment~\cite{stripes1}, where the observed radiation is sharply peaked in frequency at $\omega_{\rm JPR}$. Intuitively, the broadband photocurrent
acts like an impulsive drive to coherent Josephson plasma modes, both bulk and surface ones.
Here we focus on the surface excitations because the scenario based on the bulk ones is inconsistent with the experimental data, as we discuss in the Supplemental Materials~\cite{Note1} and further elaborate on below. In the following discussion, we investigate a semi-phenomenological model, where we assume the presence of such a broadband drive and study how it interacts with Josephson plasmons.

To further appreciate the importance of the surface excitations, we argue that the  photocurrent generation
occurs at the surface of the sample. Indeed, the frequency of the incoming photoexcitation is large so that the skin depth of how light penetrates the sample is small. From the equilibrium optical properties~\cite{stripes1}, we estimate it to be $d \approx 200\,$nm; this short electronic length scale is much shorter than any length scale associated with the collective excitations of the superconductor~\cite{PhysRevLett.105.257001,Stojchevska2011Mechanisms}. To illustrate this explicitly, we consider the bulk polariton dispersion in the vicinity of $q = 0$:
\begin{align}
    \omega_{\rm JP}^2(q_x) \approx \omega^2_{\rm JPR} + \frac{c^2q_x^2}{\epsilon_{\infty,ab}},\label{eqn:bulk_plasm}
\end{align}
where $\epsilon_{\infty,ab}$ is the high-frequency in-plane dielectric constant of the medium. Equation~\eqref{eqn:bulk_plasm} allows defining the plasmon coherence length as $l_{\rm JP} = c/(\omega_{\rm JPR}\sqrt{\epsilon_{\infty,ab}}) \sim 100\,\mu$m$\,\gg d$. Hence, from the perspective of collective modes, one indeed can consider the   photocurrent
as a surface phenomenon. We remark that in the experiment, the beam-spot size is large $d_{\rm beam} \sim 500\,\mu$m$\,\gtrsim l_{\rm JP}$ so that the pump can be approximately described as uniform along the interface.

For future reference, the electromagnetic response of layered materials, such as cuprates, is encoded into the anisotropic dielectric tensor $\hat{\varepsilon}(\omega) = \text{diag}(\varepsilon_{ab},\varepsilon_{ab},\varepsilon_{c})$. Motivated by the two-fluid model of anisotropic superconductors~\cite{Note1,two_fluid,PhysRevB.50.12831,lu2020surface,dolgirev2021periodic}, which accurately captures optical linear response of cuprate superconductors~\cite{PhysRevB.50.12831,lu2020surface}, we choose the following form:
\begin{align}
    \varepsilon_{\alpha = ab/c}(\omega) = \epsilon_{\infty,\alpha} \Big( 1 - \frac{\omega^2_{\alpha}}{\omega^2}  + \frac{i\gamma_{\alpha}}{\omega}\Big).\label{eqn_diel_func}
\end{align}
This form so far does not include the charge order -- we will return to this below. The second term in Eq.~\eqref{eqn_diel_func} describes the reactive response of the superconducting fluid so that $\varepsilon_\alpha(\omega)\sim \omega^{-2}$ for $\omega \to 0$; $\omega_{ab}$ and $\omega_c = \omega_{\rm JPR}$ are the in-plane and $c$-axis plasma frequencies. The strong anisotropy of cuprates implies $\omega_{ab}\gg \omega_c$. We also expect the in-plane plasmons to be strongly damped $\gamma_{ab}\gg \gamma_{c}$. The third term represents the normal fluid; $\gamma_{\alpha}$ is a phenomenological damping parameter, which is proportional to the corresponding normal-fluid conductivity~\cite{Note1,PhysRevB.50.12831}. Unless specified otherwise, we use the following parameters~\cite{stripes1,lu2020surface,RevModPhys.77.721}: $\omega_c = 1\,$THz, $\gamma_{c} = 0.1\,$THz, $\epsilon_{\infty, c} = 25$, and $\epsilon_{\infty, ab} = 4$. Our conclusions do not rely on the choice of $\omega_{ab}$ and $\gamma_{ab}$ as long as $\omega_{ab}\gg \omega_c$.

{\it Surface Josephson plasmons}\textemdash We begin by reviewing the properties of surface Josephson plasmons in the absence of CDW order~\cite{savel2005surface,savel2010terahertz,lu2020surface}. These are evanescent collective modes that are confined to the interface $x = 0$ [inset of Fig.~\ref{fig::Filtering_stripes}], i.e., they decay both into the air and into the sample~\cite{lu2020surface,chatterjee2021single,dolgirev2021characterizing}. Of the most physical interest for us below are the modes with the magnetic field pointing along the $y$-axis:
\begin{align}
    B_y = \begin{cases}
    B_a e^{k_a x + iq_zz  - i\omega t}, & x < 0\\
    B_m e^{-k_m x + iq_zz  - i\omega t}, & x > 0
    \end{cases}.
\end{align}
Note that because of the translational invariance along the $z$-axis, we choose the same dependence on $z$ inside and outside the sample. Here $k_a$ and $k_m$ are yet unknown wave vectors that depend on both $\omega$ and $q_z$. We find them by solving the Maxwell equations in each media:
\begin{align}
    k_a^2 =  q_z^2 - \frac{\omega^2}{c^2},\quad k_m^2 = \varepsilon_{c}\Big( \frac{q_z^2}{\varepsilon_{ab}} - \frac{\omega^2}{c^2}\Big). \label{eqn:simple_disp}
\end{align}
To describe evanescent electromagnetic waves, we choose the roots with $\text{Re}\,k_a, \text{Re}\,k_m > 0$. By matching the Fresnel boundary conditions ($E_{{\rm air},z} = E_{{\rm mat},z}$ and $B_{{\rm air},y} = B_{{\rm mat},y}$ at $x = 0$), we obtain an implicit equation on the dispersion $\omega_s(q_z)$ of the surface Josephson plasmons:
\begin{align}
    q_z = \frac{\omega_s}{c}\sqrt{\frac{\varepsilon_{ab}(\omega_s)(1 - \varepsilon_c(\omega_s)) }{1 - \varepsilon_{ab}(\omega_s)\varepsilon_c(\omega_s)}}.
\end{align}
The spectrum of these excitations is shown in Fig.~\ref{fig::Plasmon_disp}(a). We find that $\omega_s(q_z)$ quickly saturates at around $\omega_{\rm JPR}$. In other words, we expect a large density of states of these excitations near $\omega_{\rm JPR}$. It is worth pointing out that the fact that the saturation occurs near the bulk plasmon resonance,  $\omega_s(q_z)\to \omega_{\rm JPR}$ for $cq_z \gtrsim \omega_{\rm JPR}$, is a consequence of the strong anisotropy $\omega_{ab}\gg \omega_c$. For instance, in isotropic superconductors with $\omega_{ab} = \omega_c$ and $\varepsilon_{\infty, ab} = \varepsilon_{\infty, c} =1$, a similar saturation occurs but at a notably lower frequency $\omega_{\rm JPR}/\sqrt{2}$.  We finally remark that the two other surface Josephson plasmons exhibit a similar saturation but at much higher frequency.

An immediate problem we encounter is that the surface excitations lie outside the light cone and, therefore, cannot radiate. This issue is evaded by the charge order, which couples the mode with wave vector $q_z$ to the ones with wave vectors $q_z + nQ_{\rm CDW}$, where $n = 0, \pm 1, \pm 2, \dots$ and $Q = Q_{\rm CDW}$ is the ordering wave vector along the $z$-axis. As such, the entire surface plasmon dispersion becomes backfolded to the reduced Brillouin zone defined by $Q$. Some modes back fold into the light cone [Fig.~\ref{fig::Plasmon_disp}] and, therefore, they can emit photons. This insight is crucial in explaining the experiment. In essence, the stripes act as a natural diffraction grating, which then out-couples the otherwise silent surface modes, much like a nano-fabricated corrugation would be engineered~\cite{PhysRevB.52.R731,schumacher1998surface,tsiatmas2010superconducting,tsiatmas2012low,two_fluid,smith1953visible,mikhailov1998plasma,kachorovskii2012current,petrov2016plasma,petrov2017amplified}.

Here we describe the CDW order phenomenologically. Since the most dominant effect is expected to be due to the charge modulation along the $z$-axis, we neglect the fact that stripes have a non-trivial structure along the $x$- and $y$-axes. Specifically, we assume that the CDW order parameter enters through the modulation of the $c$-axis plasmon frequency:
\begin{align}
    \omega_{c} \to \omega_{c} = \omega_{\rm JPR} + \delta \omega_c(z),\quad  \delta \omega_c(z)\propto \cos(Q z).
\end{align}
This modulation, which we assume to be weak, modifies only the $c$-axis dielectric function, which we write as:
\begin{align}
    \varepsilon_c(\omega) \to \varepsilon_c(\omega) + A\cos(Q z),\label{eqn:eps_c_stripes}
\end{align}
where $A$ captures the strength of the stripes order.
For simplicity, we assume that the stripe period along the $z$-axis equals two lattice constants.
This assumption is by no means crucial but allows us to make substantial analytical progress. Finally, to properly describe the momentum mixing, one should take into account the momentum dependence of the dielectric tensor~\eqref{eqn_diel_func}. However, this dependence is expected to be nonessential at the scale of $Q$, as we show in~\cite{Note1}, where we carefully consider the surface Josephson plasmons at large momenta.

To get the properties of surface Josephson plasmons in the medium with stripes, we proceed similarly as above but now take into account the Bragg mixing of the $z$-axis momenta. Specifically, we substitute the following evanescent-wave ansatz:
\begin{align}
B_{{\rm air}, y} & = \Big[ \alpha_a e^{k_a x + i q_z z} + \beta_a e^{\tilde{k}_a x + i (q_z+Q) z}\Big] e^{-i\omega t},\label{eqn:B_air_can}\\
B_{{\rm mat}, y} & = [\alpha_m e^{-\lambda_1 x} (e^{i q_z z}  + \gamma_1 e^{i (q_z + Q) z})  \notag\\
&\qquad
+ \beta_m e^{-\lambda_2 x}(\bar{\gamma}_2 e^{i q_z z}  +  e^{i (q_z+Q) z})]e^{-i\omega t},\label{eqn:B_mat_can}
\end{align}
where $k_a(q,\omega) = \sqrt{ q^2 - \omega^2/c^2}$ and $\tilde{k}_a = k_a(q + Q,\omega)$. 
The wave vectors $\lambda_1$ and $\lambda_2$, together with parameters $\gamma_1$ and $\bar{\gamma}_2$ that encode the mentioned mixing, are known functions of $q_z$ and $\omega$, which are obtained by solving the Maxwell equations inside the sample with stripes~\cite{Note1}. The four remaining unknown coefficients $\alpha_a,\beta_a,\alpha_m,\beta_m$ are related to each other via: $B_{{\rm air}, y} = B_{{\rm mat}, y}$ and $E_{{\rm air}, z} = E_{{\rm mat}, z}$. These conditions, in turn, implicitly define the spectrum of surface plasmons through $\det {\cal M}(q_z,\omega) = 0$, where
\begin{align*}
{\cal M} \equiv
    \begin{bmatrix}
   1 + \displaystyle\frac{\varepsilon_c}{\varepsilon_c^2 - A^2} \frac{k_m^2}{\lambda_1 k_a} & \displaystyle \Big( 1 + \frac{\varepsilon_c}{\varepsilon_c^2 - A^2} \frac{k_m^2}{\lambda_2 k_a}\Big) \bar{\gamma}_2\\
    \displaystyle\Big( 1 +\frac{\varepsilon_c}{\varepsilon_c^2 - A^2}\frac{\tilde{k}_m^2}{\lambda_1 \tilde{k}_a}\Big) \gamma_1 & 1 +  \displaystyle\frac{\varepsilon_c}{\varepsilon_c^2 - A^2}\frac{\tilde{k}_m^2}{\lambda_2 \tilde{k}_a}
    \end{bmatrix}.
\end{align*}
Schematic illustration of the modified by the stripes dispersion is shown in Fig.~\ref{fig::Plasmon_disp}(b), where we reflect the back folding of high momenta surface modes into the reduced Brillouin zone.

\emph{ An impulsive drive to the surface modes}\textemdash Having established the structure of surface excitations, we turn to discuss the role of  photocurrent. 
It lives on the surface of the material and results in the emission of radiation in both the air and sample [Fig.~\ref{fig::Filtering_stripes}]. Motivated by the experiment~\cite{stripes1}, we turn to evaluate the frequency dependence of the electromagnetic field emitted into the air. We assume that 
$J_{\rm NL}$
flows along the $z$-axis so that the magnetic field is oriented along the $y$-axis. Since $J_{\rm NL}$ is confined to the surface of the superconductor, we can incorporate this drive through the generalization of the Fresnel formalism, where as above we solve Maxwell's equations inside the two media independently and then match the solutions using appropriate boundary conditions. One of them  is the continuity of the tangential component of the electric field (follows from Faraday's law): $E_{{\rm air},z} = E_{{\rm mat},z}$ at $x = 0$. The other equation is obtained from integrating the fourth Maxwell equation:
\begin{align}
    \int_{0^-}^d dx (\partial_x B_y) = \int_{0^-}^d dx  \, (\mu_0 J_{\rm NL} - i \omega \varepsilon_c(\omega) E_z/c^2).\label{eqn:BC_v1}
\end{align}
The second term in Eq.~\eqref{eqn:BC_v1} is parametrically small, $\int_{0^-}^d dx\, E_z \propto d/l_{\rm JP} \ll 1$, so we neglect it and obtain:
\begin{align}
B_{{\rm air},z} = B_{{\rm mat},z} - \mu_0 d J_{\rm NL}.\label{eqn:Fresnel_mod}
\end{align}

Given the possibility of having a non-zero Umklapp photocurrent~\cite{dolgirev2021Umklapp}, we write Eq.~\eqref{eqn:shift_current_gen} as
\begin{align}
    J_{\rm NL}(z,\omega) = J_0 + J_U\cos (Q_{\rm CDW} z).~\label{eqn:J_shift}
\end{align}
This insight that one can have $J_U \neq 0$ is essential for understanding the experimental data, as we show below. Using the form~\eqref{eqn:J_shift}, we turn to compute the spectrum of outgoing radiation. To this end, we employ the same ansatz as in Eqs.~\eqref{eqn:B_air_can}-\eqref{eqn:B_mat_can}, except we now specialize on $q_z = 0$. By invoking the derived boundary conditions, we arrive at:
\begin{gather}
      \begin{bmatrix}
       \alpha_a\\
       \beta_a
    \end{bmatrix} =
   \left( \begin{bmatrix}
       1 & \bar{\gamma}_2\\
       \gamma_1 & 1
    \end{bmatrix} {\cal M}^{-1} 
-    \begin{bmatrix}
       1 & 0\\
       0 & 1
    \end{bmatrix}
    \right)
    \begin{bmatrix}
       \mu_0 d J_0\\
       \mu_0 d J_U
    \end{bmatrix}.\label{eqn:main}
\end{gather}
The coefficient $\alpha_a$, in turn, gives the amplitude of the emitted into the air radiation $f(\omega)$ -- see Fig.~\ref{fig::Filtering_stripes}.

In the absence of the Umklapp component $J_U = 0$, we find that $f(\omega)$ displays a double peak structure, which comes from the splitting of the bulk JP resonance, encoded in the prefactors $\varepsilon_c/(\varepsilon_c^2 - A^2)$. It is worth pointing out that since the surface excitations exhibit saturation around $\omega_{\rm JPR}$, i.e., at the bulk resonance, it is not entirely clear that this splitting comes from the bulk rather than the surface. To resolve this question we consider in \cite{Note1} isotropic superconductors, where the surface excitations are well separated from the bulk ones, and confirm that the double-peak structure originates from the bulk. We also further elaborate in \cite{Note1} on the asymptotic behavior of $f(\omega)$ at both small and large frequencies. This double-peak splitting was not observed in the experiment~\cite{stripes1}. In addition, the resulting spectral function is too broad for realistic parameters to explain the experimental data.

Most remarkably, provided $J_U\neq 0$, we find the emission spectrum $f(\omega)$ becomes sharply peaked at around $\omega_{\rm JPR}$. This peak originates from the fact that the Umklapp component now resonantly drives the surface plasmons at wave vectors around $q_z = Q$; due to the back folding into the light cone, these plasmons can now radiate photons, as illustrated in Fig.~\ref{fig::Plasmon_disp}(b). We further elaborate on the surface origin of this effect in \cite{Note1}. This result that the Umklapp  photocurrent 
can give rise to a sharp emission provides a natural interpretation of the experimental data.

\begin{figure}[t!]
\centering
\includegraphics[width=1\linewidth]{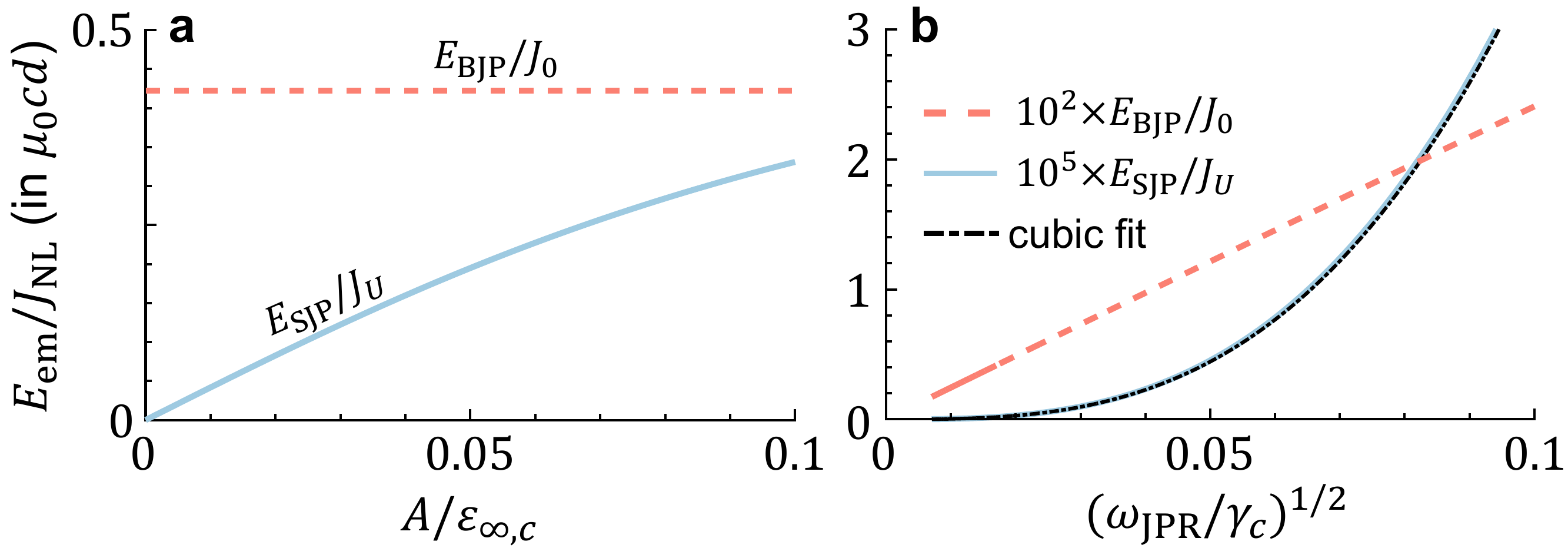} 
\caption{ Differences between the bulk and surface scenarios. (a) Emission amplitude $E_{\rm SJP}/J_{U}$ from the surface plasmon scales linearly with the CDW amplitude $A$, Eq.~\eqref{eqn:eps_c_stripes}, while the emission in the bulk case $E_{\rm BJP}/J_{0}$ is independent of $A$. (b) For $\omega_{\rm JPR}\ll\gamma_c$, we find that $E_{\rm SJP}\sim(\omega_{\rm JPR}/\gamma_c)^{3/2}$ for the surface scenario and $E_{\rm BJP}\sim(\omega_{\rm JPR}/\gamma_c)^{1/2}$ for the bulk one. 
}
\label{fig::Scaling}
\end{figure}

\emph{Bulk plasmon scenario}\textemdash We now comment on the role of bulk polaritons. Since they are also being impulsively driven by $J_0\neq 0$, Eq.~\eqref{eqn:main}, they can produce characteristic radiation peaked in frequency near $\omega_{\rm JPR}$. While we cannot completely rule out the bulk scenario as a possible alternative interpretation of the experiment~\cite{stripes1,Note1}, there are two good reasons why the observed emission is dominated by the surface Josephson plasmons. The first one is that the resulting bulk emission spectrum turns out to be too broad for realistic parameters to explain the experiment~\cite{Note1}. In contrast, in the surface scenario, the Umklapp photocurrent 
resonantly drives the surface modes resulting in sharp radiation without any fine-tuning of model parameters. 
The second argument invokes the experimental phenomenology that the observed radiation linewidth correlates with the out-of-plane CDW coherence length (once inversion symmetry is broken). This fact is inconsistent with the bulk scenario because the far-field reflectivity, which entirely determines the bulk emission properties, is known to be insensitive to the stripes~\cite{stripes1,Rajasekaran.2018}. At the same time, the observed phenomenology can naturally be explained within the surface scenario since the corresponding emission is acutely sensitive to the CDW order (for an additional discussion, see~\cite{Note1}) -- see also Fig.~\ref{fig::Scaling}(a).

Finally, to further distinguish the two scenarios from each other, we theoretically predict the scaling of each response with temperature $T$. As $T$ approaches $T_c$ from below the JPR softens to zero. As shown in Fig.~\ref{fig::Scaling}, we find that the outgoing emission exhibits distinct behaviors for the two scenarios when $\omega_{\rm JPR}\ll \gamma_c$:
\begin{align*}
    E_{\rm BJP} \propto J_0 \Big( \frac{\omega_{\rm JPR}}{\gamma_c}\Big)^{1/2} \text{ and } E_{\rm SJP} \propto J_U A \Big( \frac{\omega_{\rm JPR}}{\gamma_c}\Big)^{3/2}.
\end{align*}
One important implication of this result is that the emission amplitude in the surface scenario $E_{\rm SJP}$ is expected to become suppressed compared to the bulk one $E_{\rm BJP}$ for $T\approx T_c$: $E_{\rm SJP}/E_{\rm BJP}\propto \omega_{\rm JPR}/\gamma_c$. Since according to Eq.~\eqref{eqn:main} both scenarios contribute to the outgoing radiation, we conclude that the surface scenario dominates at lower temperatures, while the bulk one might play a role in the immediate vicinity of $T_c$.

\emph{Conclusion}\textemdash 
To summarize, our interpretation of the observed terahertz emission in striped superconductors consists of two related arguments. The first one is the downconversion of optical fields by fermionic quasiparticles from high-frequency high-intensity pump pulse to low-frequency regular and Umklapp photocurrents. 
Given that the emission is observed only in the striped phase, we expect that this downconversion arises from the CDW order~\cite{dolgirev2021Umklapp}. The second argument is that photocurrents
impulsively drive collective modes of the sample. In particular, the Umklapp component resonantly couples to the surface Josephson plasmons, in turn resulting in spectrally sharp radiation at $\omega_{\rm JPR}$  and, thus, explaining the experiment~\cite{stripes1}. In the electrodynamics of striped superconductors, both stripe and superconducting orders combine and play significant roles.

For outlook, an intriguing open question is the microscopic origin of the photocurrents
and whether it can be related to a pair density wave~\cite{Berg_2009}.
The Umklapp photocurrent
is interesting on its own, even without superconductivity, and might be relevant, for instance, in moir\'{e} systems~\cite{dolgirev2021Umklapp,chaudhary2021shift}.

\begin{acknowledgments}
We thank M.D.~Lukin, M.~Mitrano, I.~Esterlis, N. Leitao, and P.~Narang for fruitful discussions. P.E.D., M.H.M., and E.D. were supported by AFOSR-MURI: Photonic Quantum Matter award FA95501610323, DARPA DRINQS, and the ARO grant ``Control of Many-Body States Using Strong Coherent Light-Matter Coupling in Terahertz Cavities''. J.B.C. is supported by the Quantum Science Center (QSC), a National Quantum Information Science Research Center of the U.S. Department of Energy (DOE), and by the Harvard Quantum Initiative. J.B.C. also acknowledges hospitality from the Max Planck Institute for Structure and Dynamics of Matter (MPSD, Hamburg), and ETH Zürich Institute for Theoretical Physics. D.N., M.B., M.F., and A.C. acknowledge support from the European Research Council under the European Union’s Seventh Framework Programme (FP7/2007-2013)/ERC Grant Agreement No. 319286 (QMAC), the Deutsche Forschungsgemeinschaft (DFG, German Research Foundation) via the excellence cluster ‘The Hamburg Centre for Ultrafast Imaging’ (EXC 1074 – project ID 194651731), and the priority program SFB925 (project ID 170620586). 
\end{acknowledgments}

\footnotetext[1]{See Supplemental Material for additional theoretical details. The discussion there includes five sections: i) plane waves in the medium with stripes; ii) bulk polariton scenario; iii) surface Josephson plasmons at large momenta; iv) bulk vs surface contributions to the spectrum of outgoing radiation; and v) reflection from a medium with Bragg mixing. The Supplemental Material includes Refs.~\cite{stripes1,Rajasekaran.2018,Pitarke06,Lu2020,Alpeggiani13,Tinkham}. }

\footnotetext[2]{Said differently, nonlinear terms quadratic in the optical field and linear in plasmon fields are forbidden in materials with inversion symmetry.}

\bibliography{Sc_lib}

\cleardoublepage
\onecolumngrid
\begin{large}
\begin{center}
\textbf{Supplemental Material to ``Theory for Anomalous Terahertz Emission in Striped Cuprate Superconductors''}
\end{center}
\end{large}
\vspace{0.5cm}
\twocolumngrid

\beginsupplement

\section{Plane waves in the medium with stripes}
Here we analyse plane waves in the sample with stripes. Because of the translational symmetry breaking along the $z$-axis, the structure of free harmonics contains the mixing between $q_z$ and $q_z + Q$ wave vectors; at the same time, along the $x$-axis, we can substitute an evanescent wave ansatz:
\begin{align}
    B_{{\rm mat},y} = (\alpha e^{iq_z z} + \beta e^{i(q_z + Q)z}) e^{-\lambda x-i\omega t},
\end{align}
where $\lambda(\omega,q_z)$ is yet unknown wave vector that depends on both $\omega$ and $q_z$; $\alpha$ and $\beta$ are two coefficients that encode the strength of the mentioned mixing. Substituting this ansatz into the fourth Maxwell equation with the modified dielectric function, Eq.~\eqref{eqn:eps_c_stripes} of the main text, 
we obtain the electric field in the sample: 
\begin{align*}
    E_{{\rm mat},x} &  = \frac{c^2}{\omega \varepsilon_{ab} } \Big[  q_z\alpha e^{iq_z z} + (q_z + Q) \beta e^{i(q_z + Q)z}\Big] e^{-\lambda x-i\omega t},\\
    E_{{\rm mat},z} & = -\frac{i c^2\lambda \varepsilon_c }{\omega (\varepsilon_c^2 - A^2) } \Big[ \Big(\alpha - \frac{A}{\varepsilon_c} \beta \Big) e^{iq_z z} \notag\\
    & \qquad\qquad\qquad\quad
    + \Big(\beta - \frac{A}{\varepsilon_c} \alpha \Big)  e^{i(q_z + Q)z}\Big] e^{-\lambda x-i\omega t}.
\end{align*}
Plugging this result into the third Maxwell equation, we get the following secular equation:
\begin{align}
\displaystyle\begin{bmatrix*}
    k_m^2(q_z,\omega) - \lambda^2 & A\lambda^2/\varepsilon_c \\
    A\lambda^2/\varepsilon_c & k_m^2(q_z+Q,\omega) - \lambda^2
\end{bmatrix*} \begin{bmatrix}
   \displaystyle \alpha\\
   \displaystyle \beta 
\end{bmatrix} = \begin{bmatrix*}
   \displaystyle 0\\
   \displaystyle 0
\end{bmatrix*}, \label{eqn:eigenvalue_stripes}
\end{align}
where $k_m(q,\omega)$ is determined implicitly through
\begin{align}
    \displaystyle \frac{q^2}{\varepsilon_{ab}} - \frac{k_m^2(q,\omega ) \varepsilon_c }{\varepsilon_c^2 - A^2 } = \frac{\omega^2}{c^2}.
\end{align}
For consistency, when solving this equation, we choose the root with $\text{Re}\, k_m(q,\omega) > 0$.
There are four eigenvalues of Eq.~\eqref{eqn:eigenvalue_stripes}:
\begin{align*}
    \lambda = \displaystyle \pm \sqrt{ \frac{k_m^2 + \tilde{k}_m^2 \pm \sqrt{ (k_m^2 + \tilde{k}_m^2)^2 - 4k_m^2\tilde{k}_m^2 \Big( 1 - \displaystyle \frac{A^2}{\varepsilon_c^2} \Big)  } }{ 2 \Big( 1 -\displaystyle \frac{A^2}{\varepsilon_c^2} \Big)}  },
\end{align*}
where we defined $\tilde{k}_m = k_m(q_z + Q,\omega)$. We select the two of them with $\text{Re}\,\lambda_1,\,\text{Re}\,\lambda_2 >0$ that describe waves decaying into the sample. We choose these roots such that in the limit $A\to 0$, we get $\lambda_1 \approx k_m$ and $\lambda_2 \approx \tilde{k}_m$. Equation~\eqref{eqn:eigenvalue_stripes} also fixes the ratio between the amplitudes $\alpha$ and $\beta$:
\begin{align}
    \bar{\gamma}_\lambda \equiv \gamma_{\lambda}^{-1} \equiv \frac{\alpha}{\beta} =   \frac{\lambda^2}{\lambda^2 - k_m^2} \frac{A}{\varepsilon_c } = \frac{\lambda^2 - \tilde{k}_m^2}{\lambda^2}
    \frac{\varepsilon_c}{A}.
\end{align}

\section{Bulk Polariton Scenario}
\label{sec:bulk}

In this section, we consider a simplified phenomenological model, where we assume the presence of a broadband photocurrent
, Eq.~\eqref{eqn:shift_current_gen} of the main text, but we neglect stripes and do not take into account how they modify the dielectric properties of the sample. This is a bit artificial because the photocurrent generation 
itself is expected to occur due to the stripes~\cite{stripes1}. Nevertheless, this approach allows us to separate the response due to the bulk Josephson plasmons from the surface modes since the latter remain silent because they lie outside the light cone, as shown in Fig.~\ref{fig::Plasmon_disp}(a) of the main text. In other words, the emission we compute below comes entirely due to the bulk excitations.

\begin{figure}[t!]
\centering
\includegraphics[width=1\linewidth]{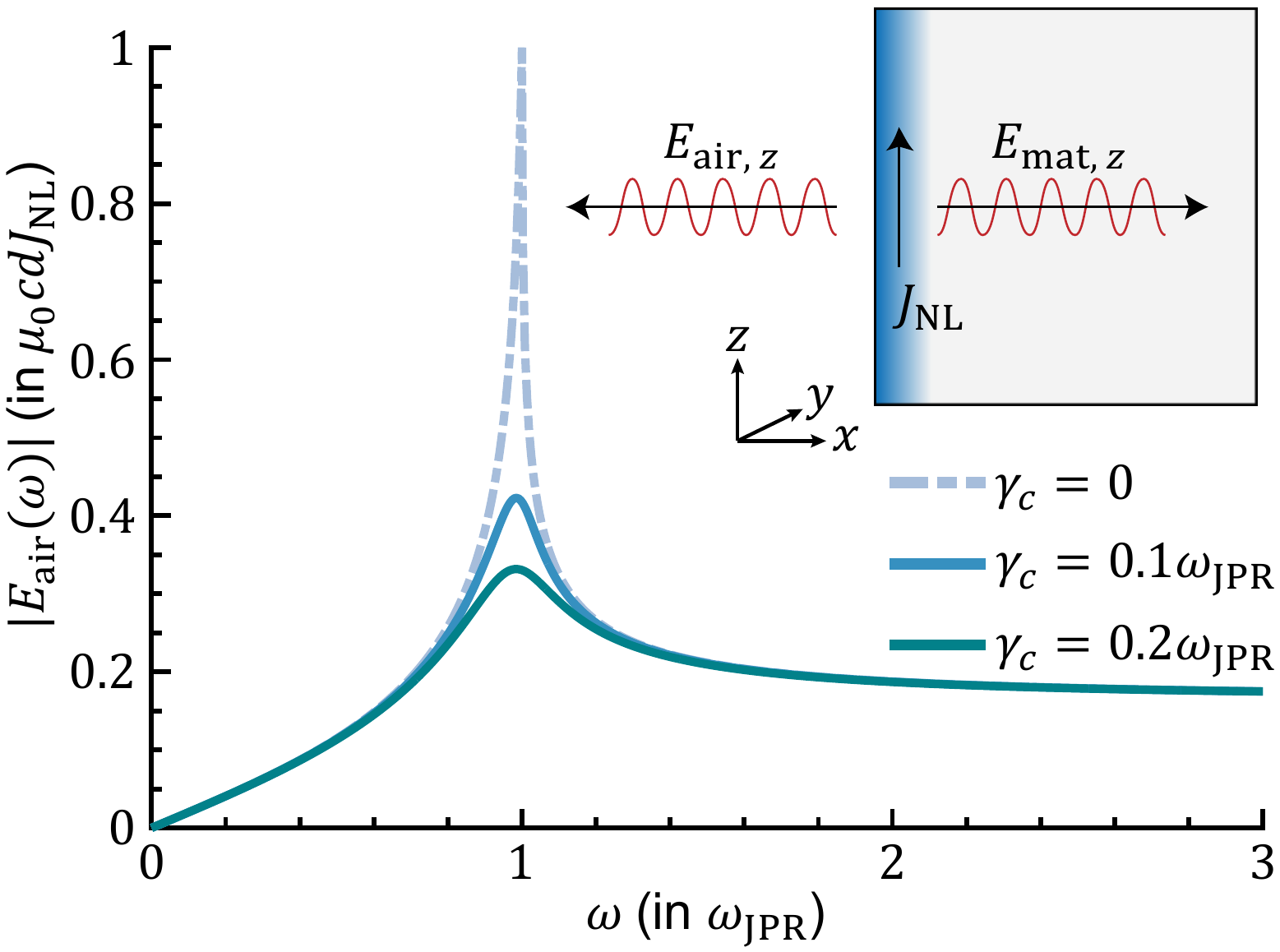} 
\caption{ Bulk polariton scenario. While the photocurrent 
in Eq.~\eqref{eqn:shift_current_gen} does not depend on frequency $\omega$, for $\omega \lesssim \omega_{\rm JPR}$, we find that the amplitude of the radiation into the air displays a peak at $\omega_{\rm JPR}$. This peak, though, is smeared out for realistic values of the quasiparticle damping $\gamma_{c}$. The inset illustrates that the photocurrent 
serves as a drive that emits radiation into the air and the superconductor. }
\label{fig::Filtering_no_stripes}
\end{figure}

To get the emission properties in this model, we first solve the Maxwell equations in each media. In the air, we have an outgoing plane wave, which we write as:
\begin{equation}
    B_{{\rm air}, y} = B_a e^{-i\omega x/c - i\omega t},\, E_{{\rm air}, z} =  c  B_a e^{-i\omega x/c - i\omega t},
\end{equation}
where $B_a(\omega)$ encodes the amplitude of this plane wave. In the superconductor, the solution reads:
\begin{align}
    B_{{\rm mat}, y} = B_m e^{iq_m x - i\omega t},
\end{align}
where $q_m(\omega)$ is yet an unknown wave vector that depends on $\omega$, and $B_m(\omega)$ is the amplitude of this medium harmonic. From the fourth Maxwell equation, we obtain the electric field in the sample:
\begin{align}
    E_{{\rm mat}, z} = -\frac{q_m c^2}{\omega \varepsilon_c} B_m e^{iq_m x - i\omega t}.
\end{align}
By plugging this into the third Maxwell equation, we get: $q_m^2 = \omega^2 \varepsilon_c(\omega)/c^2$. Now, using the generalized Fresnel boundary conditions, derived in the main text, we finally obtain:
\begin{align}
    B_a(\omega) = - \frac{\mu_0 d J_{\rm NL}}{1 + \sqrt{\varepsilon_c(\omega)}} = f(\omega)\mu_0 d J_{\rm NL} .\label{eqn:filt_no_stripes}
\end{align}
This is the central result of this section, which we turn to discuss in detail.

Figure~\ref{fig::Filtering_no_stripes} shows the amplitude $f(\omega)$ of the emitted into the air radiation as a function of frequency $\omega$. We find that even though the photocurrent 
is flat for $\omega \lesssim \omega_{\rm JPR}$, $f(\omega)$ is peaked at $\omega_{\rm JPR}$, i.e., the superconductor acts a low-frequency filter to the featureless drive $J_{\rm NL}(\omega)$. The reason for this peaked behavior is the enhanced due to the van Hove singularity density of states of bulk $c$-axis Josephson plasmons at $\omega = \omega_{\rm JPR}$. We now turn to discuss the dependence of $f(\omega)$ in more detail. First, at large frequencies $\omega \gtrsim \omega_{\rm JPR}$, where the photon dispersion inside the sample (bulk polariton) becomes approximately linear, the drive can emit light into both the air and the material. The fraction of how much it emits into the air is entirely determined by the relative speed of light $c/c_m = \sqrt{\varepsilon_{\infty,ab}}$ in the air to that in the superconductor. For instance, in case $c = c_m$, half of the radiation would go into the air and the other half into the sample. Second, for small frequencies $\omega\to 0$, we find that $f(\omega)$ becomes suppressed and, in particular, $f(0 ) = 0$. The reason for this is that fundamentally, the superconductor can easily screen dc current, not allowing a low-frequency electric field to develop.

We comment that the peak at $\omega_{\rm JPR}$ can be smeared out by finite $\gamma_{c}$. For realistic parameters, this smearing effect is profound, and the resulting emission spectrum is too broad to explain the experiment~\cite{stripes1}. In contrast, the surface scenario provides sharp emission even if $\gamma_c$ is notable essentially because the Umklapp photocurrent 
{\it resonantly} drives the surface modes.

We turn to discuss an additional argument why the bulk scenario is inconsistent with the experimental phenomenology. In the experiment~\cite{stripes1}, the emission was detected in two LBCO samples corresponding to two different dopings. These samples have stripes with different relative strengths and out-of-plane coherence lengths, and the resulting emission was more intense and much narrower in frequency for the sample with stronger, more coherent stripes.  The issue with the bulk scenario, however, is that the far-field reflectivity, which entirely determines the emission properties associated with the bulk modes, is known to be insensitive to the stripes~\cite{stripes1,Rajasekaran.2018}. Therefore, the bulk plasmons are expected to produce similar radiation in the two samples, with the same lineshapes independent of the stripe correlation length and, thus, disagreeing with the experiment. In contrast, the emission from the surface Josephson plasmons is acutely sensitive to the CDW order, further justifying our expectation that the observed phenomenon comes from the surface modes.

\section{Surface Josephson plasmons at large momenta}
\label{appendix:two-fluid}

In the main text, we considered bulk $c$-axis Josephson plasmons with flat dispersion, as encoded in momentum-independent electric permittivity~\eqref{eqn_diel_func}. If one is interested in small momenta close to the light cone $q\simeq\omega/c$, then this approach is well-justified. However, we also encountered in the main text the situation of Bragg mixing of surface modes with small momenta to those that have large momenta $q \simeq Q_{\rm CDW}$. For these latter modes, it might be essential to consider the momentum dependence of $\hat{\varepsilon}(\omega,q)$, as we do in this section. On the phenomenological level, the leading correction to the flat $c$-axis dispersion comes from the effects of quasiparticle compressibility, which, in turn, sets a very large Thomas-Fermi momentum scale $q_{\rm TF}$. We expect that in cuprates $Q_{\rm CDW} \lesssim q_{\rm TF} \lesssim 2\pi/a_0$, implying that: i) $Q_{\rm CDW}$ is still small enough so that one can essentially disregard momentum dependence of $\hat{\varepsilon}(\omega,q)$ for $q \lesssim Q_{\rm CDW}$; ii) the lattice reciprocal momentum $2\pi/a_0$ is large, and effects of compressibility cannot be ignored at such a momentum scale~\cite{Pitarke06}. In the literature, the dispersions of the usual surface plasmons and the surface Josephson plasmons in stacked 2D metals and anisotropic superconductors have been discussed in various geometries, including infinite half-space, thin film~\cite{Lu2020}, and spherical particles~\cite{Alpeggiani13}. In our theory, we go beyond previous treatments of surface Josephson plasmons by including the effects of quasiparticle compressibility, which allows us to capture the dispersion of surface excitations at large momenta of the order of $Q_{\rm CDW}$.

We describe the electric response of superconductors within the two-fluid model~\cite{Tinkham}, where the total electric current $\bm J = \bm J_n + \bm J_s$ is represented as a sum of the quasiparticle contribution $\bm J_n$ and the one due to the superflow $\bm J_s$. For the normal component, we write Ohm's law but take into account the effects of electrochemical potential $\delta \mu = \delta \rho/\chi$, where $ \delta \rho$ represents the electric charge density and $\chi$ is the compressibility:
\begin{align}
\bm J_n = \hat{\sigma} (\bm E - \nabla \delta \mu).
\end{align}
For simplicity, we assume that the quasiparticle conductivity tensor $\hat{\sigma} = \text{diag}(\sigma_{ab},\sigma_{ab},\sigma_c)$ is  frequency and momentum independent. For the superconducting part, we write London's equation:
\begin{align}
    \partial_t \bm J_s =  \hat{\Lambda}(\bm E - \nabla \delta \mu).
\end{align}
Within the BCS theory, the tensor $\hat{\Lambda} = \text{diag}(\Lambda_{ab},\Lambda_{ab},\Lambda_c)$ is proportional to $|\psi|^2$, where $\psi$ is the superconducting order parameter. We also have the continuity equation:
\begin{align}
    \partial_t \delta \rho + \nabla \cdot \bm J = 0.\label{eqn:cnt_v1}
\end{align}
Using $\varepsilon_0 \nabla \cdot \bm E = \delta\rho$ and $\nabla \times \bm B = \mu_0 (\bm  J_n + \bm J_s) + \partial_t \bm E/c^2$, we evaluate the dielectric tensor of the material:
\begin{align}
    \hat{\varepsilon}_{\alpha\beta}(\bm q,\omega)
    & =\varepsilon_\alpha \delta_{\alpha \beta} + (\varepsilon_\alpha - 1 )\frac{q_\alpha q_\beta}{q_{\rm TF}^2},\label{eqn:eps_anisotropic}
\end{align}
where 
\begin{align}
    \varepsilon_\alpha =  1 - \frac{\omega^2_\alpha}{\omega^2}  + \frac{i{\gamma}_\alpha} {\omega}. \label{eqn:eps_simplif}
\end{align}
Here we used that $\Lambda_\alpha = \varepsilon_0\omega_\alpha$ and $\sigma_\alpha =\varepsilon_0\gamma_\alpha$ and defined the Thomas-Fermi momentum as $q_{\rm TF}^2 = \chi/\varepsilon_0$. Equation~\eqref{eqn:eps_anisotropic} implies that the $c$-axis Josephson plasmon dispersion is no longer flat: $\omega_c^2(q_z) = \omega_{\rm JPR}^2 (1 + q_z^2/q_{\rm TF}^2)$. In the limit $q_{\rm TF} \to \infty$, we reproduce the result in Eq.~\eqref{eqn_diel_func}, up to unimportant for this discussion anisotropy factors.

We turn to investigate plane waves in this dispersive medium and plug in the usual ansatz:
\begin{gather}
    \bm B = B_y \hat{y}\, e^{i q_x x + iq_z z - i\omega t},\\
    \bm E = (E_x \hat{x} + E_z \hat{z}) e^{i q_x x + iq_z z - i\omega t}.
\end{gather}
Using the third and fourth Maxwell equations, we obtain an implicit equation on $q_x(q_z,\omega)$: $\det {\cal M}(q_z,\omega) = 0$, where ${\cal M}$ is a $2\times 2$ matrix given by
\begin{align*}\displaystyle  
\begin{bmatrix}
\displaystyle       \varepsilon_{ab} -\frac{q_z^2 c^2}{\omega^2} + \frac{(\varepsilon_{ab} - 1) q_x^2}{q_{\rm TF}^2}  & \quad
\displaystyle
 q_x q_z\Big[
 \frac{ \varepsilon_{ab} - 1 }{q_{\rm TF}^2}+ \frac{ c^2}{\omega^2} \Big]\\ \quad
        \displaystyle q_x q_z\Big[
 \frac{ \varepsilon_{c} - 1 }{q_{\rm TF}^2}+ \frac{ c^2}{\omega^2} \Big]
 &        \displaystyle
 \varepsilon_{c} - \frac{q_x^2 c^2}{\omega^2} + \frac{(\varepsilon_{c} - 1) q_z^2}{q_{\rm TF}^2} 
    \end{bmatrix}. 
\end{align*}
This secular equation is only bi-quadratic in $q_x$. By solving it, we find four roots, out of which only two describe waves that decay into the sample ($q_x = i k_m$, so that $\text{Re}\, k_m^{(1)},\text{Re}\, k_m^{(2)} > 0$). We choose these roots such that in the limit $q_{\rm TF}\to \infty$, the harmonic $k_m^{(1)}$ reduces to the result~\eqref{eqn:simple_disp} of the main text:
\begin{align}
    [k_m^{(1)}]^2 = \displaystyle \varepsilon_{c}\Big( \frac{q_z^2}{\varepsilon_{ab}} - \frac{\omega^2}{c^2}\Big) \Big[ 1  -\frac{(\varepsilon_{ab} - \varepsilon_c)^2 q_z^2}{\varepsilon_{c}\varepsilon_{ab}^2 q_{\rm TF}^2 } + \dots\Big].
\end{align}
Notably, as $q_{\rm TF}\to \infty$, the second root $k_m^{(2)}\sim q_{\rm TF}$ diverges, which essentially means that the second harmonic can be disregarded, as we elaborate below.

Our subsequent task is to study the surface collective excitations. Since for given $q_z$ and $\omega$, we have two distinct harmonics in the sample, we write the following ansatz:
\begin{align}
    B_z = e^{iq_z z-i\omega t} \begin{cases}
    B_0 e^{k_a x}, & x< 0\\
    B_1 e^{-k_m^{(1)} x} + B_2 e^{-k_m^{(2)} x}, & x > 0
    \end{cases}.\label{eqn:B_gen_large_q}
\end{align}
Amplitudes $B_0$, $B_1$, and $B_2$ are related to each through appropriate boundary conditions, which we turn to derive. From the first Maxwell equation, we obtain:
\begin{align}
\varepsilon_0(E_{{\rm mat}, x} - E_{{\rm air}, x}) = \delta\rho_{\rm 2D},
\end{align}
where $\delta\rho_{\rm 2D}$ is the two-dimensional surface charge density, which we assume to be nonzero. Faraday's law gives:
\begin{align}
    E_{{\rm mat},z} = E_{{\rm air},z}.
\end{align}
From Ampere's law, we obtain that the two-dimensional surface current flows along the $z$-axis, $\bm J_{\rm 2D}\parallel \hat{z}$, which, in turn, gives:
\begin{align}
    B_{{\rm mat},y} - B_{{\rm air},y} = \mu_0 J_{{\rm 2D},z}.
\end{align}
By integrating the continuity equation~\eqref{eqn:cnt_v1} near the surface, we finally get:
\begin{align}
    \partial_t \delta \rho_{\rm 2D}  + \partial_z J_{{\rm 2D},z} + J_{{\rm mat},x} = 0.\label{eqn:cnt_v2}
\end{align}
We note that the third term represents the 3-dimensional current density, reflecting the fact that the surface charge can leak into the bulk of the sample. To obtain a closed system of equations, we use the medium equations and find:
\begin{gather*}
   \mu_0  J_{{\rm mat},x} = -\frac{i
   \omega(\varepsilon_{ab} - 1) }{c^2}  \Big[E_{{\rm mat},x}   - \frac{\partial_x\partial_\alpha E_{{\rm mat},\alpha} }{q_{\rm TF}^2}\Big],\\
   \mu_0 J_{{\rm 2D},z} = - \frac{ \omega q_z}{c^2 q_{\rm TF}^2}  (\varepsilon_c - 1)  [ E_{{\rm mat},x} - E_{{\rm air},x}].
\end{gather*}
These medium relations, together with Eqs.~\eqref{eqn:B_gen_large_q}-\eqref{eqn:cnt_v2}, form a closed set of equations, which allows us to obtain the spectrum of the surface Josephson plasmons at arbitrary momenta.

\begin{figure}[t!]
\centering
\includegraphics[width=1\linewidth]{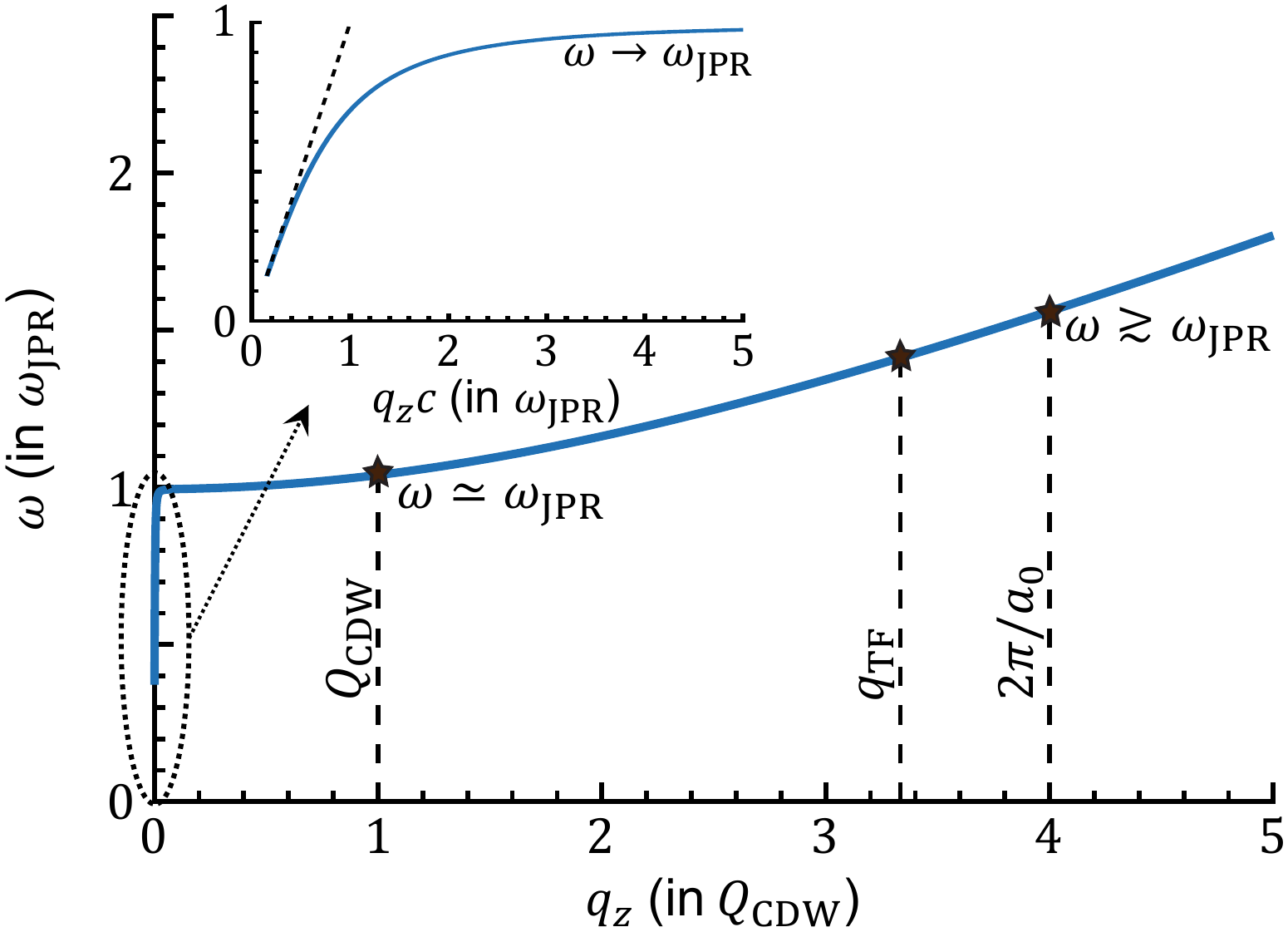}
\caption{ Dispersion $\omega_s(q_z)$ of the surface excitations for large momenta $q_z\simeq q_{\rm TF}$, where the effects of quasiparticle compressibility become notable.  In contrast to the behavior at low momenta near the light cone (inset), see also Fig.~\ref{fig::Plasmon_disp}(a), where $\omega_s(q_z)$ displays a quick saturation near $\omega_{\rm JPR}$, now $\omega_s(q_z)$ is no longer flat. In cuprates, we expect $Q_{\rm CDW}\lesssim q_{\rm TF}\lesssim 2\pi/a_0$. In this case, the stripes momentum $Q_{\rm CDW}$ is still quite small, $\omega_s (Q_{\rm CDW}) \simeq \omega_{\rm JPR}$, and the effects of compressibility are unimportant at this scale. At the same time, the lattice momentum $2\pi/a_0$ is large resulting in $\omega_s (2\pi/a_0)$ being substantially different from $\omega_{\rm JPR}$.
}
\label{fig::TF}
\end{figure}

While the above set of equations can be directly (numerically) solved, here we aim at understanding the effects of quasiparticle compressibility perturbatively, to the leading order in $q_{\rm TF}^{-2}$. To this end, one can neglect the second strongly damped harmonic $k_m^{(2)}$ in Eq.~\eqref{eqn:B_gen_large_q}
\begin{align*}
    B_z = e^{iq_z z-i\omega t} \begin{cases}
    B_a e^{k_a x}, & x< 0\\
    B_m e^{-k_m^{(1)} x}, & x > 0
    \end{cases}
\end{align*}
and solve the following boundary problem:
\begin{gather*}
    E_{{\rm mat},z} = E_{{\rm air},z},\\
    B_{{\rm mat},y} - B_{{\rm air},y} = - \frac{ \omega q_z (\varepsilon_c - 1)}{c^2 q_{\rm TF}^2}    [ E_{{\rm mat},x} - E_{{\rm air},x}].
\end{gather*}
We remark that this perturbative approach satisfies the continuity equation~\eqref{eqn:cnt_v2} to the leading order in $q_{\rm TF}^{-2}$. In other words, the role of the harmonic $k_m^{(2)}$ is to satisfy the charge conservation at higher orders. For the spectrum of surface Josephson plasmons, we obtain the following approximate equation:
\begin{align}
    \frac{1}{k_a} \Big[ 1 + \frac{q_z^2(\varepsilon_c - 1)}{q_{\rm TF}^2 } \Big] + \frac{\varepsilon_c}{k_m^{(1)}} \Big[ 1 + \frac{q_z^2(\varepsilon_c - 1)}{\varepsilon_c q_{\rm TF}^2 } \Big] = 0.
\end{align}
Figure~\ref{fig::TF} shows the resulting dispersion $\omega_s(q_z)$. While at low momenta, $q_z \ll q_{\rm TF}$, we reproduce the same behavior as in Fig.~\ref{fig::Plasmon_disp}(a), i.e., $\omega_s(q_z)$ shows a quick saturation near $\omega_{\rm JPR}$, at larger momenta, $\omega_s(q_z)$ becomes dispersive. In particular, for momenta $q_z \gtrsim q_{\rm TF}$, we expect that $\omega(q_z)$ will be substantially different from $\omega_{\rm JPR}$. The result in Fig.~\ref{fig::TF} implies that for the interpretation we offer in the main text to be consistent, in the experiment, we should have $Q_{\rm CDW} \lesssim q_{\rm TF}\lesssim 2\pi/a_0$ so that $\omega(Q_{\rm CDW})\approx \omega_{\rm JPR}$. This latter requirement is essential from the perspective of back folding of the modes with momenta $q_z \simeq Q_{\rm CDW}$, which should have frequency close to $\omega_{\rm JPR}$.

\section{Bulk vs surface contributions to the spectrum of outgoing radiation}
\label{appendix:isotropic}

So far, we primarily studied strongly anisotropic superconductors with $\omega_{ab}\gg \omega_c$. A unique feature of such materials is that the surface modes saturate at the frequency immediately below the bulk Josephson plasmon resonance $\omega_{\rm JPR}$. This makes distinguishing whether the features in the spectral function of outgoing radiation arise from the surface or bulk modes difficult. To resolve this question, here we solve the problem for isotropic superconductors, where the surface plasmon frequency $\omega_{\rm JPR}/\sqrt{2}$ is well separated from the bulk.

\begin{figure}[t!]
\centering
\includegraphics[width=1\linewidth]{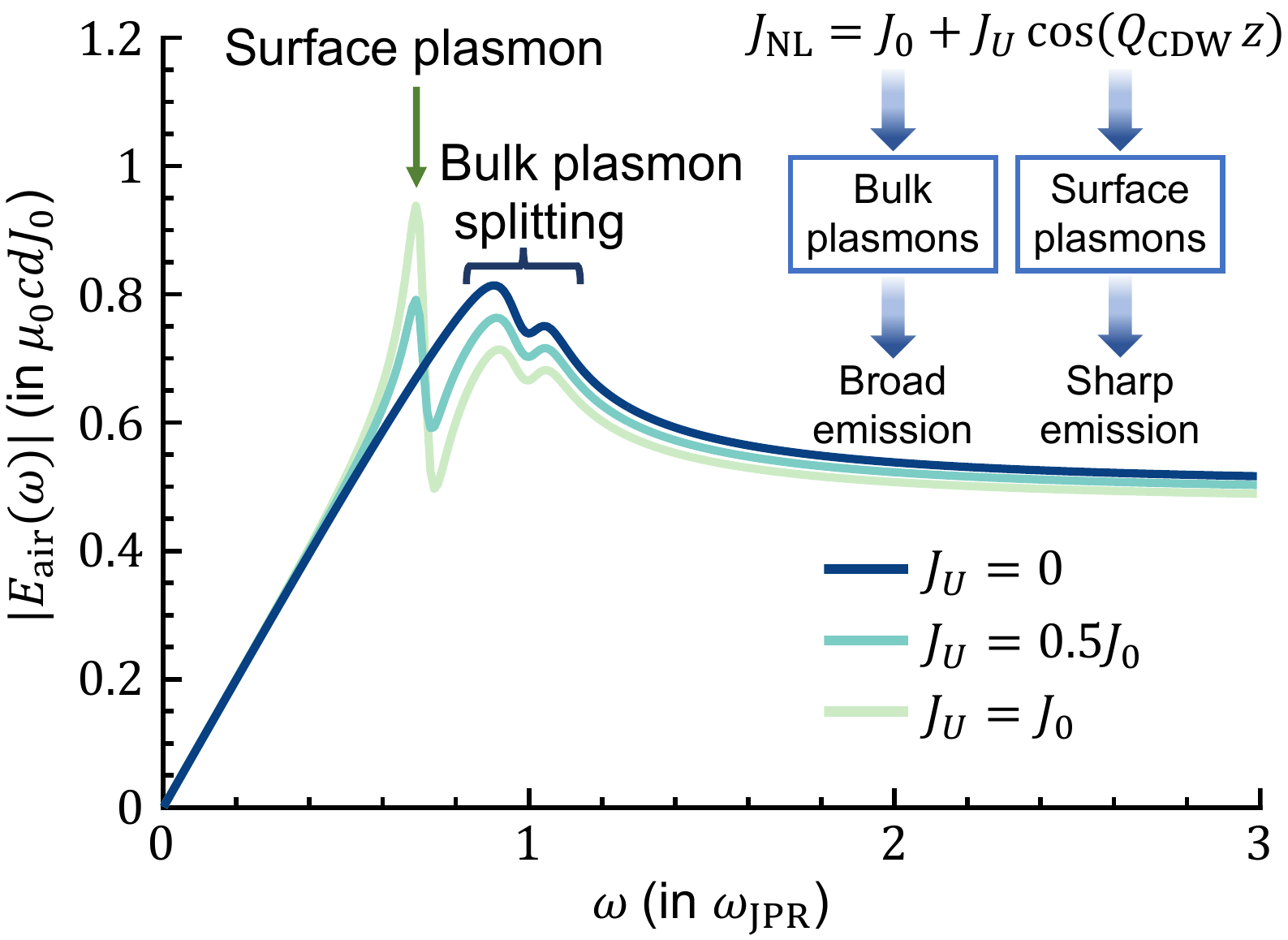} 
\caption{Emission spectrum $f(\omega)$ in a system with $\omega_{ab} = \omega_c = 1\,$THz, $\gamma_{ab} = \gamma_{c} = 0.1\,$THz, $\varepsilon_{\infty,ab} = \varepsilon_{\infty,c} =1$, $A = 0.1$. In such isotropic superconductors, bulk and surface resonances have notably different energies. This allows us to conclude that the homogeneous component $J_0$ of the photocurrent 
drives the bulk plasmons only, while the Umklapp one 
$J_U$ affects the surface plasmons, in turn, providing a sharp resonant structure of $f(\omega)$.
}
\label{fig::Filt_iso}
\end{figure}

Figure~\ref{fig::Filt_iso} shows the emission spectral function $f(\omega)$ in isotropic superconductors. For $J_U = 0$, $f(\omega)$ displays a double-peak structure near the bulk resonance: the dip that appears at $\omega_{\rm JPR}$ corresponds to the bulk plasmon splitting in the striped phase due to the hybridization between zero and finite momentum bulk modes. Importantly, once $J_U \neq 0$, $f(\omega)$ exhibits an additional sharp peak at the surface resonance. This confirms both that the sharp resonance in Fig.~\ref{fig::Filtering_stripes} is due to the surface Josephson plasmons and that surface Josephson plasmons can only be resonantly excited by a nonzero Umklapp photocurrent. 
We remark that the bulk features are not severely renormalized by $J_U$. We conclude that the zero momentum photocurrent $J_0$ drives bulk modes and leads to a broad emission, while the Umklapp photocurrent $J_U$ primarily drives surface excitations and leads to a sharp peak at the surface plasma resonance -- see the inset in Fig.~\ref{fig::Filt_iso} for a summary.

\section{ Reflection from a medium with Bragg mixing }
\label{app:refl}

\begin{figure}[t!]
\centering
\includegraphics[width=1\linewidth]{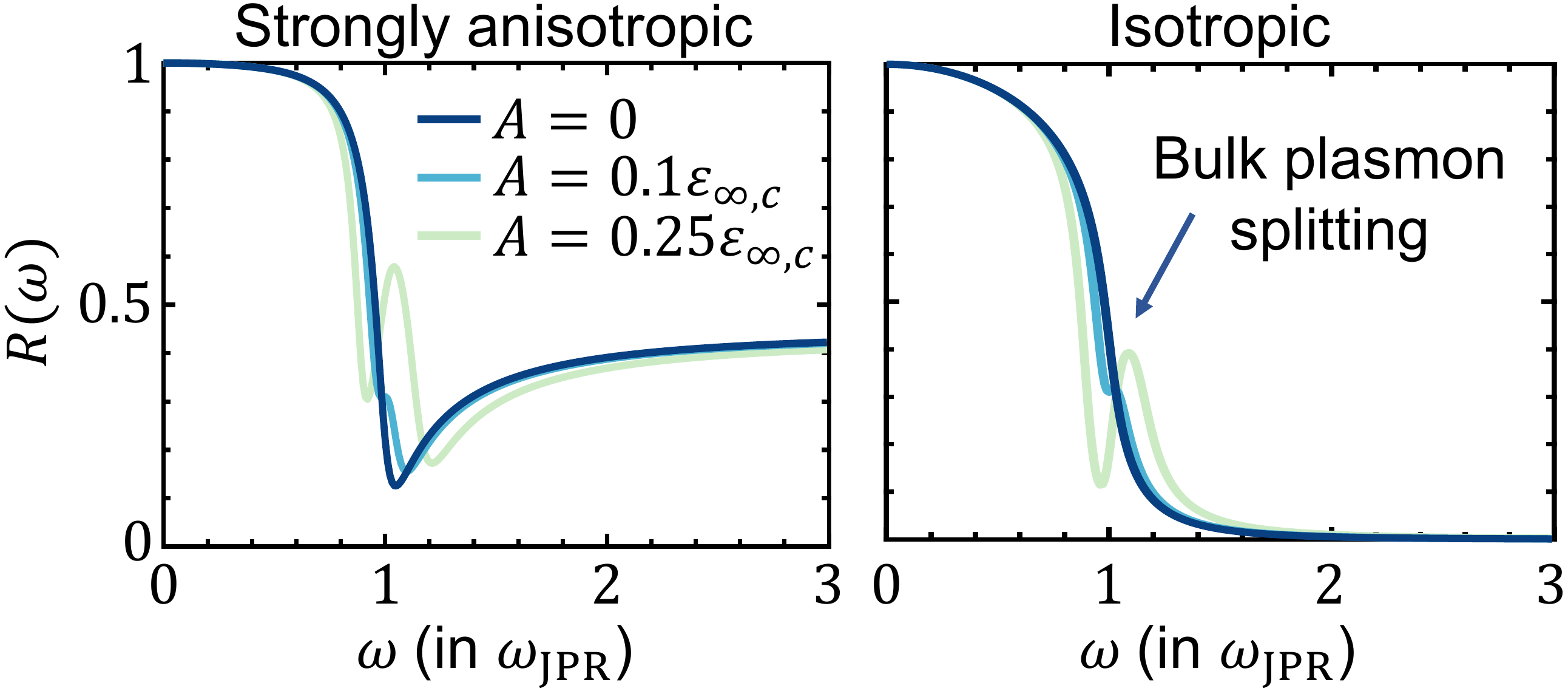}
\caption{ Reflectivity $R(\omega)$ in anisotropic (left) and isotropic (right) superconductors with stripes. Provided the amplitude $A$ of the CDW order is nonzero, there occur new features in $R(\omega)$ near the bulk plasmon resonance $\omega_{\rm JPR}$. At the same time, as indicated in the right panel, no features occur at the surface resonance, implying that the far-field photons are coupled only to the bulk plasmons. Parameters in the left (right) panel are the same as in Fig.~\ref{fig::Filtering_stripes} (Fig.~\ref{fig::Filt_iso}).
}
\label{fig::Refl}
\end{figure}

Here we compute the reflection coefficient $r_p$ of the medium with stripes. Following the discussion in the main text, cf. Eqs.~\eqref{eqn:B_air_can}-\eqref{eqn:B_mat_can}, we write the magnetic field as ($q_z= 0$):
\begin{align}
    & B_{{\rm air},y}  =  [\alpha_a ( e^{- k_a x} + 
r_p e^{k_a x }) + \beta_a e^{\tilde{k}_a x + i Q z}] e^{-i\omega t},\\
& B_{{\rm mat},y}  =  [\alpha_m e^{-\lambda_1 x} (1  + \gamma_1 e^{i Q z})  \notag\\
&
\qquad\qquad\qquad\qquad
+ \beta_m e^{-\lambda_2 x}(\bar{\gamma}_2   +  e^{i Q z})] e^{-i\omega t}.
\end{align}
Now, the coefficients $\alpha_a,\beta_a,\alpha_m,\beta_m$ are related to each through the appropriate Fresnel boundary conditions, which in turn give an implicit equation on $r_p$: $\det {\cal N} = 0$, where 
\begin{widetext}
\begin{align}
    {\cal N} = \begin{bmatrix}
   \displaystyle \frac{1}{r_p + 1} + \displaystyle\frac{\varepsilon_c}{\varepsilon_c^2 - A^2} \frac{k_m^2}{\lambda_1 k_a} \displaystyle \frac{1}{r_p - 1}& \displaystyle \Big( \displaystyle \frac{1}{r_p + 1} + \frac{\varepsilon_c}{\varepsilon_c^2 - A^2} \frac{k_m^2}{\lambda_2 k_a}\displaystyle \frac{1}{r_p - 1}\Big) \bar{\gamma}_2\\
    \displaystyle\Big( 1 +\frac{\varepsilon_c}{\varepsilon_c^2 - A^2}\frac{\tilde{k}_m^2}{\lambda_1 \tilde{k}_a}\Big) \gamma_1 & 1 +  \displaystyle\frac{\varepsilon_c}{\varepsilon_c^2 - A^2}\frac{\tilde{k}_m^2}{\lambda_2 \tilde{k}_a}
    \end{bmatrix}.
\end{align}
\end{widetext}
For $A=0$, we reproduce the familiar expression:
\begin{align}
    r_p = \frac{\varepsilon_c k_a - k_m}{\varepsilon_c k_a + k_m} .
\end{align}

We compare the usual Josephson plasma edge appearing in the reflectivity spectra of a layered superconductor without stripes to reflectivity spectra renormalized by the stripes in Fig.~\ref{fig::Refl}~(left). We find that for $A\neq 0$, the Bragg mixing between zero momentum, $q_z = 0$, and finite momentum, $q_z = Q_{\rm CDW} $, gives rise to a plasma edge splitting. As shown in Fig.~\ref{fig::Refl}~(right), where following the preceding section, we consider isotropic superconductors and confirm that the new features in Fig.~\ref{fig::Refl}~(left) come from the bulk. In other words, far-field measurements, even in the presence of the CDW order, are sensitive to the bulk and not the surface excitations. However, up to relatively large stripe amplitudes, $A \sim 0.1 \epsilon_{\infty,c}$, the new features appear negligible and can be easily missed within the experimental precision. This is consistent with reflectivity experiments in stripes which have reported only a single plasma edge. For this reason, in our calculations for terahertz emission, we use the value of $A = 0.1 \epsilon_{\infty,c}$ as a reasonable upper limit to the stripe amplitude. 

Notably, we find that the coupling to finite momentum modes can drastically enhance terahertz emission even though reflectivity probes are essentially insensitive to these modes up to large values of the stripe strength $A$.

\end{document}